%% file: main.tex
\documentclass{vgtc}                          

\ifpdf
  \pdfoutput=1\relax                   
  \usepackage{graphicx}                
  \DeclareGraphicsExtensions{.pdf,.png,.jpg,.jpeg} 
\else
  \usepackage{graphicx}                
  \DeclareGraphicsExtensions{.eps}     
\fi%

\graphicspath{{figures/}{pictures/}{images/}{./}} 

\usepackage{microtype}                 
\PassOptionsToPackage{warn}{textcomp}  
\usepackage{textcomp}                  
\usepackage{mathptmx}                  
\usepackage{times}                     
\usepackage{cite}                      
\usepackage{tabu}                      
\usepackage{booktabs} 
\usepackage{xcolor}

\usepackage{paralist}
\usepackage{soul}
\usepackage{censor}
\usepackage{ulem}
\usepackage[font={scriptsize,sf}]{caption}
\onlineid{0}

\vgtccategory{Research}

\vgtcinsertpkg

\title{A Visual Analytics System for Water Distribution System Optimization
\vspace{-5pt}
}

\author{
Yiran Li\thanks{e-mail: \{ranli, enmusabandesu, tfujiwara, fjloge, klma\}@ucdavis.edu} %
\and Erin Musabandesu\footnotemark[1] %
\and Takanori Fujiwara\footnotemark[1] %
\and Frank J. Loge\footnotemark[1] %
\and Kwan-Liu Ma\footnotemark[1]
}
\affiliation{\vspace{-7pt} \scriptsize University of California, Davis \vspace{-13pt}}

\abstract{\vspace{-1pt}

The optimization of water distribution systems (WDSs) is vital to minimize energy costs required for their operations.
A principal approach taken by researchers is identifying an optimal scheme for water pump controls through examining computational simulations of WDSs.
However, due to a large number of possible control combinations and the complexity of WDS simulations, it remains non-trivial to identify the best pump controls by reviewing the simulation results.
To address this problem, we design a visual analytics system that helps understand relationships between simulation inputs and outputs towards better optimization.
Our system incorporates interpretable machine learning as well as multiple linked visualizations to capture essential input-output relationships from complex WDS simulations.
We demonstrate our system's effectiveness through a practical case study and evaluate its usability through expert reviews. 
Our results show that our system can lessen the burden of analysis and assist in determining optimal operating schemes.

} 

\nocopyrightspace

\begin{document}



\maketitle

\input{text_files/1introduction.tex}

\input{text_files/2related_work.tex}
\input{text_files/3analysis_questions.tex}
\input{text_files/4visualization_design.tex}

\input{text_files/5case_study.tex}

\input{text_files/6evaluation.tex}

\input{text_files/7discussions.tex}
\input{text_files/8conclusion.tex}


\acknowledgments{
This research is supported in part by the U.S. National Science Foundation through grant IIS-1741536.
We acknowledge funding from the California Energy Commission Electric Program Investment Charge, agreement number EPC-16-062. 
We would also like to thank Moulton Niguel Water District for their partnership on this project and other research regarding the water-energy nexus.

This document was prepared as a result of work sponsored by the California Energy Commission. It does not necessarily represent the views of the Energy Commission, its employees, or the State of California. The Energy Commission, the State of California, its employees, contractors, and subcontractors make no warranty, express or implied, and assume no legal liability for the information in this document; nor does any party represent that the use of this information will not infringe upon privately owned rights. This report has not been approved or disapproved by the Energy Commission nor has the Energy Commission passed upon the accuracy of the information in this report.
}


\bibliographystyle{abbrv-doi}

\bibliography{main}
\end{document}

%% file: text_files/1introduction.tex
\vspace{-1pt}
\section{Introduction}
\vspace{-1pt}
Water distribution systems (WDSs) are complex networks that deliver water from sources to customers. 
They include pipes, junctions, tanks, pumps, and valves. 
WDSs span almost one million miles in the United States and represent the vast majority of physical infrastructure for water supply~\cite{epa_2019}. 
The optimal WDS design and operation should minimize the capital investment and energy costs while meeting hydraulic constraints, fulfilling water demands, and satisfying pressure requirements~\cite{mays_2000}. 

Our research focuses on minimizing energy costs in existing WDSs. 
Energy costs often vary by time of day, with customers paying higher prices when the demand for energy is the highest.
WDSs can manage the timing of energy use to lower energy costs by changing pump operations. 
They can further alter pump operations by pumping to an elevated water storage facility during low pricing periods and use gravity to serve customers during peak price times.

Hydraulic modeling software such as EPANET~\cite{epanet} can run hydraulic simulations efficiently to facilitate the optimization of WDS operations. 
By examining the simulation results of different pump operation controls, researchers can better understand how these controls affect energy costs. 
However, as the number of pumps increases, running simulations of all control combinations to find optimal operating schemes can become intractable. 
EPANET only allows the user to view the results of a single simulation at a time; consequently, comparing the results of multiple simulations is challenging. 
If many control combinations and simulation outcomes are tested, the analysis must be done externally, and researchers may find it hard to digest and discover insights from such large data. 

WDS pump operations can be more efficiently optimized using a metaheuristic simulation optimization procedure~\cite{Lost2017}. 
These methods use hydraulic simulations to evaluate how different operations perform, testing out different control combinations according to a given algorithm to reduce energy cost within predefined constraints.
However, these methods act as a black box~\cite{april2003practical}.
Little information is extracted on which control combinations are tested or how important reliability parameters vary based on different control sets.

In this work, we develop a visual analytics system that helps WDS researchers understand how simulation results vary as pump operations are changed and extract actionable insights to achieve better and more transparent optimization of WDS operations. 
The analysis core of our system utilizes interpretable machine learning (ML)---specifically, we employ a decision tree---to identify underlying relationships between input variables and outputs from complex hydraulic simulations.
Our system provides multiple visualizations that are designed to observe the simulation data from multiple aspects, such as the similarities of input variables and temporal changes of WDS status.
By effectively integrating these visualizations and an interpretable ML method, our system supports the exploration of multiple sets of simulation data, extraction of rules to optimize pumping operations, and verification of the extracted rules. 
We demonstrate the usability of this system through a case study on a WDS widely used for the analysis and evaluate the effectiveness of our system through domain experts' feedback. 
Also, we provide a discussion on the extensibility of our approach to other applications.

%% file: text_files/2related_work.tex
\begin{figure*}
 \centering
 \includegraphics[width=\linewidth]{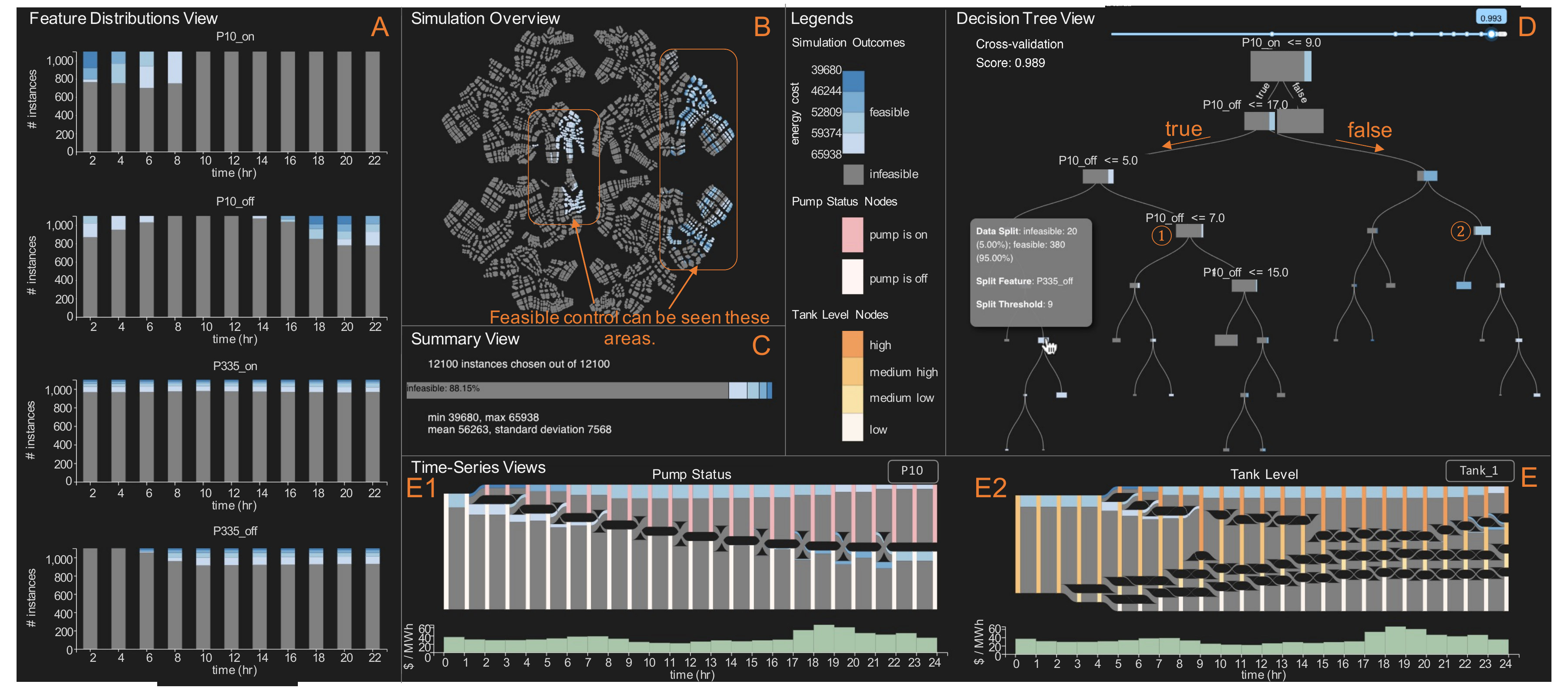}
 \caption{
 The interface of the visual analytics system for WDS optimization. Here, the NET3 simulations are analyzed. The system is composed of (A) the feature distributions view, (B) simulation overview, (C) summary view, (D) decision tree view, and (E) time-series views.
 }
 \label{fig:teaser}
\end{figure*}

\vspace{-2pt}
\section{Background and Related Work}
In this section, we describe background and related works in WDS optimization methods and visualization of simulation ensembles.

\vspace{-3pt}
\subsection{Optimization on WDS Energy Cost}
\vspace{-2pt}

There has been abundant research on optimizing operations of WDSs to minimize energy consumption~\cite{Awe_2019,PECCI2015937,w12061732}.
Pump operating costs, which constitute the largest expenditure of WDSs~\cite{VanZyl2004}, are the main focus of optimization.
Explicit pump scheduling, which controls when pumps operate, is the most widely-used formulation to define this optimization problem~\cite{Lost2017}. 
In this paper, we use the time when pumps are scheduled to be turned on or off as the control variables. 

Hydraulic modeling software, such as EPANET~\cite{epanet}, can be used to test out different pump control schemes to reduce energy costs while maintaining reliability. 
Once a model detailing the WDS is constructed, EPANET generates time-series simulations of water pressures, flows, and storage based on different operating procedures~\cite{epanet}. 
However, hydraulic modeling software typically only shows one set of simulation results at a time. 
This limitation makes comparing many different control schemes challenging. 
Metaheuristic simulation-optimization algorithms can optimize WDS operations more efficiently than the manual trial and error using EPANET. 
However, 
they do not support the comparison of many different simulation results.  
It is difficult to understand which combinations of controls have been tested by the algorithm and how the results of these simulations vary. 
Also, due to many potential control combinations and constraints, it is challenging for the algorithms to find more optimal solutions, and they may get stuck in a local minimum~\cite{Maier2014}.

Our system allows researchers to explore the results of multiple simulations in a more efficient way than using hydraulic simulations alone and provides more information about the simulation results than  metaheuristic simulation-optimization implementations.
The insights from this system can also be used to inform metaheuristic optimization strategies, allowing for more targeted optimization. 

\vspace{-3pt}
\subsection{Visualization of Simulation Ensembles}
\vspace{-2pt}

Our work visualizes a large number of WDS simulation results
and aims to extract insights from the distribution of simulation inputs and outputs.
This shares the same goal with visual analytics of simulation ensembles~\cite{Wang2018}.  
Simulation ensembles are a collection of spatio-temporal outputs generated via computational simulations~\cite{Potter2009}.
Visual analytics of simulation ensembles can be seen in various applications, such as weather prediction~\cite{Sanyal2010}, high-energy physics~\cite{Hao2016}, fluid dynamics~\cite{Hummel2013}, and particle hydrodynamics~\cite{Waser2010}.
Throughout a comprehensive survey, Wang et~al.~\cite{Wang2018} identified two common approaches of ensemble data visualization: (1) data aggregation with statistical summaries and (2) visual composition, such as juxtaposition of multiple information. 
Although WDS simulations usually do not have a strong focus on spatial information unlike simulation ensembles, our visual analytics system also takes these two approaches together to analyze large sets of complex WDS simulation data. 
When compared with existing research on simulation ensembles, where simple statistical summaries (e.g., mean and probability distributions) are often employed, our approach utilizes interpretable ML to extract a summary of relationships between large numbers of simulation inputs and outputs. 
Also, to the best of our knowledge, this paper is the first research addressing a
visual analytics of multiple simulation outcomes of WDSs.

%% file: text_files/3analysis_questions.tex
\vspace{-2pt}
\section{Design Requirements}
\label{sec:DRs}
\vspace{-2pt}

With our team's expertise, which includes visual analytics and WDS optimization, and inspired by the analytics tasks defined for ensemble visualization~\cite{Wang2018},
we identified design requirements (DRs) of a visual analytics system for WDS optimization.

\textbf{DR1}: Provide an \emph{overview} of the distribution of outcomes over changes in input variables. 
To convey general information of how input controls affect outcomes, the overview should clearly show which range of each input variable 
leads to infeasible and feasible controls as well as more optimal controls in case for feasible controls.
An informative overview can guide researchers to filter the data to review more detailed information. 

\textbf{DR2}: Support \emph{filtering} a subgroup of simulations to help find more insights specific to researcher's interest. 
For example, the researcher may want to know why simulations within a certain range of some input variable contain a large number of infeasible controls.

\textbf{DR3}: Extract underlying \emph{rules} of simulations to associate inputs and
simulation outcomes. 
The extracted rules can provide directly actionable insights to researchers for their next iteration of simulations. 
Furthermore, the rules can help understand the behavior of the complex simulation model.

\textbf{DR4}: Support \emph{verifications} and \emph{explanations} of insights learned through the analysis. 
Researchers are also interested in how the extracted rules can be explained and verified by looking at the related information. 
Providing details of the WDS status during the simulation assists understanding the reasoning behind the rules.

%% file: text_files/4visualization_design.tex
\vspace{-3pt}
\section{Visualization Design}
\label{sec:vis}
\vspace{-2pt}

As shown in~\autoref{fig:teaser}, we design a visual analytics system composed of five views: (A) the \emph{feature distributions view}, (B) \emph{simulation overview}, (C) \emph{summary view}, (D) \emph{decision tree view}, and (E) \emph{time-series views}. 
While (A--C) are designed for \textbf{DR1}, (D) and (E) are for \textbf{DR3} and \textbf{DR4}, respectively.
To support \textbf{DR2}, each view is fully linked and is updated based on a user's selection.
Each view provides fundamental interactions, such as brushing, lasso selection, and window scrolling, as demonstrated in the supplementary video.

\begin{table}
    \caption{ \vspace{-1pt}Data structure used in the analysis of WDS simulations.
    \centering
    \includegraphics[width=0.9\linewidth]{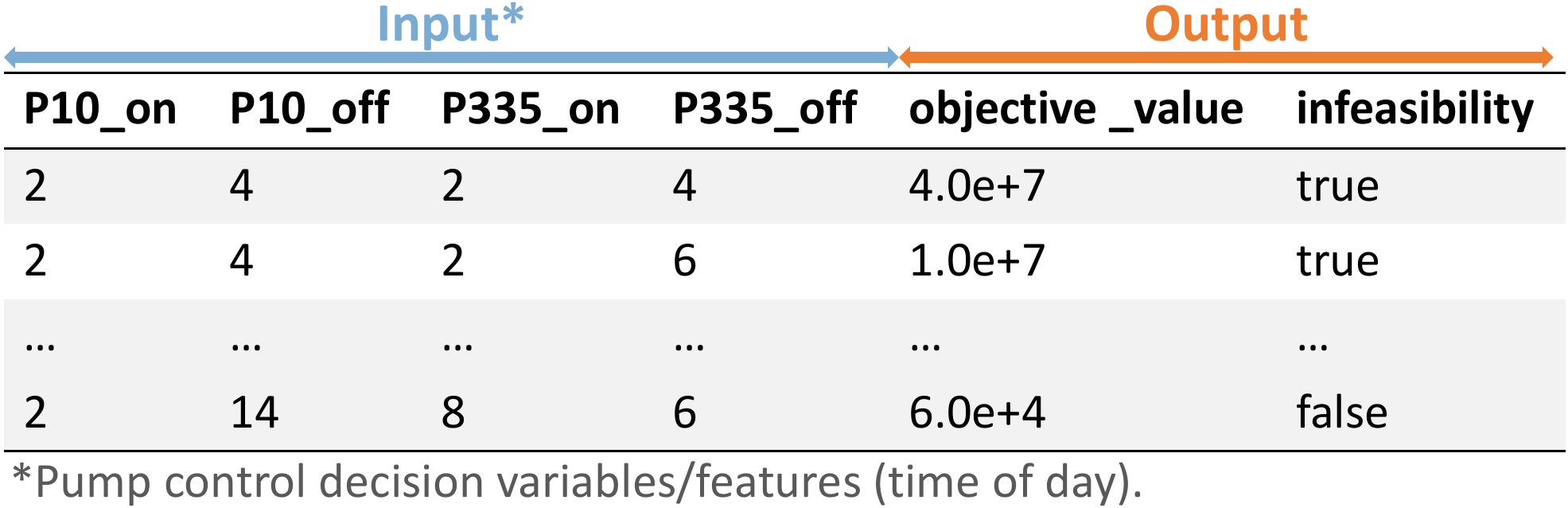}
    }
    \vspace{-6pt}
    \label{fig:data}
\end{table}

\vspace{-4pt}
\subsection{Data Description}
\label{sec:data}
\vspace{-2pt}

\autoref{fig:data} describes the structure of data we extracted from WDS simulations. 
The pump operation controls are formulated as time points when the pump is turned on or off and treated as inputs to the simulation model.
The objective values are formulated as the energy costs obtained via a simulation plus penalties for violations of hydraulic constraints.
These constraints are designed to maintain WDS reliability and include parameters, such as the minimum pressure as well as the minimum, maximum, and average tank storage levels.
If there is no violation, the penalty equals zero.
In this case, we categorize the corresponding input as a \emph{feasible} solution.
If there is one or more violations, a large penalty term is added to the objective value, and the value becomes incomparably larger when compared with the feasible solutions. 
We categorize an input that violates a constraint as an \emph{infeasible} solution. 

Below, we describe each view by applying the system to the simulation results of a real-world WDS, called NET3~\cite{epanet}. 
The NET3 model consists of two controllable pumps, named \texttt{P10} and \texttt{P335}.
We collected simulation results by setting time of turning on and off each pump with a grid sampling interval of two hours of the day.
For example, \texttt{P10\_on} (timing of turning on \texttt{P10}) is in $\{2, 4, \cdots, 22\}$ and a simulation can be set to \texttt{P10\_on}\,=\,2, \texttt{P10\_off}\,=\,6, \texttt{P335\_on}\,=\,4, and \texttt{P335\_off}\,=\,6.
This collection of simulations contains 12,100 simulation instances.
Even though we used the NET3 model for demonstration purposes, we have tested our system with more complex WDSs, such as the RW4 system (4 pumps) and Richmond skeleton system (7 pumps)~\cite{VanZyl2004}.

\vspace{-2pt}
\subsection{Feature Distributions View}
\vspace{-2pt}

This view uses a stacked histogram to present how the simulation outcomes are distributed with changes in each individual pump control.
For each pump, the simulation requires two variables (or features)---the timing of when a pump is turned \textit{on} and \textit{off}. 
Thus, \autoref{fig:teaser}A consists of four histograms (2 pumps $\times$ 2 features).
Each histogram's $x$- and $y$-axes represent an input feature value (i.e., time) and frequency (i.e., the number of simulation instances).
As we used grid sampling on a consistent interval, at an initial state, we have the same number of simulation instances for each feature value. 
This number can be changed based on filtering performed in this or other views. 
The color of each bar encodes the simulation output, as shown in legends in \autoref{fig:teaser}. 
We assign a gray color for simulation instances resulting in infeasible solutions.
Feasible solutions are colored with a blue sequential colormap based on their resultant objective values (energy costs) as we are interested in finding optimal pump controls. 

\vspace{-2pt}
\subsection{Simulation Overview}
\vspace{-2pt}

While the feature distribution view provides an overview of relationships between simulation inputs and outputs with a focus on each individual feature value, 
the simulation overview (\autoref{fig:teaser}B) depicts the same relationships with the similarity of each simulation instance's input feature values.   
From high-dimensional data (e.g., 4 features in \autoref{fig:teaser}), we embed the instance similarities into a 2D space with a dimensionality reduction method (specifically, we use t-SNE~\cite{tsne}).
Each point represents one simulation instance and the distance between each point represents their feature value similarity.
We use the same color encoding as the feature distributions view.
From this view, users can see a general pattern of the relationships, such as a cluster of instances with feasible solutions.

\vspace{-2pt}
\subsection{Summary View}
\vspace{-2pt}

The summary view (\autoref{fig:teaser}C) shows the selected instances' statistical summary, including the number of selected instances, the distribution of simulation outcomes, and statistical measures (e.g., mean) of objective values with feasible solutions. 

\vspace{-2pt}
\subsection{Decision Tree View}
\vspace{-2pt}

From a collection of simulation instances, we want to extract a rule that informs us which kinds of combinations of input features likely result feasible or infeasible solutions to help inform further optimization.
For this purpose, a collection of simulations can be considered as high-dimensional data with labels (i.e., feasible or infeasible). 
To identify features that highly influence the differences of instances with different labels, several existing visual analytics approaches utilize interpretable ML methods, such as simple neural networks~\cite{knittel2021visual}, linear discriminant analysis~\cite{wang2017linear}, and contrastive learning~\cite{fujiwara2019supporting,zhang2021visual}. 
While these approaches can extract simple rules, such as values of Features X and Y highly contribute the differences, such rules are insufficient for our case. 
As each pump control affects other pumps' feasible controls, we need to extract rules that consider a combination of multiple pump controls. 
For example, when \texttt{P10} is turned on before 9 AM and then turned off after 5 PM, simulations might judge inputs as feasible solutions no matter when turning on \texttt{P335}, and vice versa.
To extract such interpretable rules considering multiple conditions, we utilize a decision tree (DT) algorithm~\cite{DT}. 

The system applies a DT algorithm to the whole simulation data and then uses cost-complexity pruning~\cite{DT} to make the extracted DT simple and easy-to-understand. 
When pruning more nodes, while the tree less precisely describes the differences between feasible and infeasible solutions, the tree becomes simpler (i.e., less nodes and less depth); thus, easier to interpret.
A balance between the preciseness and simplicity can be controlled with a slider placed at the top of \autoref{fig:teaser}D, where the selectable prediction accuracy scores are marked with circles. 
These accuracy scores also present the trustworthiness of the DT representation.
Furthermore, an average cross-validation score besides the slider shows the DT's generalizability.

Similar to conventional DT visualization, we show the splitting criterion at each tree node (e.g., \texttt{P10\_on}\,$\leq$\,9.0). 
Left and right edges correspond to conditions of `true' and `false', respectively.
We further encode essential information in each node.
The size of each node's rectangle represents the number of instances that are split by the node.
Also, the rectangle shows the proportion of feasible and infeasible solutions with widths of inner rectangles. 
While we use the same color encoding with the feature distributions view, the feasible proportion is colored based on the mean of feasible solutions' objective values.
For example, the nodes annotated with \textcircled{\small 1} and \textcircled{\small 2} contain mostly infeasible and feasible solutions, respectively. 
By clicking a node, the user can select instances belonging to the node.

\vspace{-2pt}
\subsection{Time-Series Views}
\vspace{-2pt}

WDS simulation software such as EPANET can generate time-series data during each simulation, including pump statuses and tank levels at every hour.
We visualize this data with Sankey diagrams~\cite{sankey}. 
In the Sankey diagrams, nodes at each hour represent pump statuses at that hour, i.e., whether it is turned on or off (\autoref{fig:teaser}-E1) or water levels of a tank from high to low (\autoref{fig:teaser}-E2).
Relative vertical positions of nodes are also used to convey the same information.
The links represent the number of instances shifting from one status/level to another status/level at the next hour.
Each link also shows proportions of infeasible (grey) and feasible (blue) instances.
A user can select a pump and a tank based on their interest from a drop down menu located at the top of each diagram.
Pump statuses and tank levels are color-encoded with pink and orange sequential colormaps, respectively.
A bar chart placed below each Sankey diagram shows the average energy cost during each hour.
By correlating the width and types of links across different hours and the heights of the average energy costs, users can understand how feasibility, pump and tank statuses, and energy costs dynamically influence each other.

%% file: text_files/5case_study.tex
\vspace{-2pt}
\section{Case Study}
\vspace{-2pt}

\begin{figure}[tb]
    \centering
    \includegraphics[width=\linewidth]{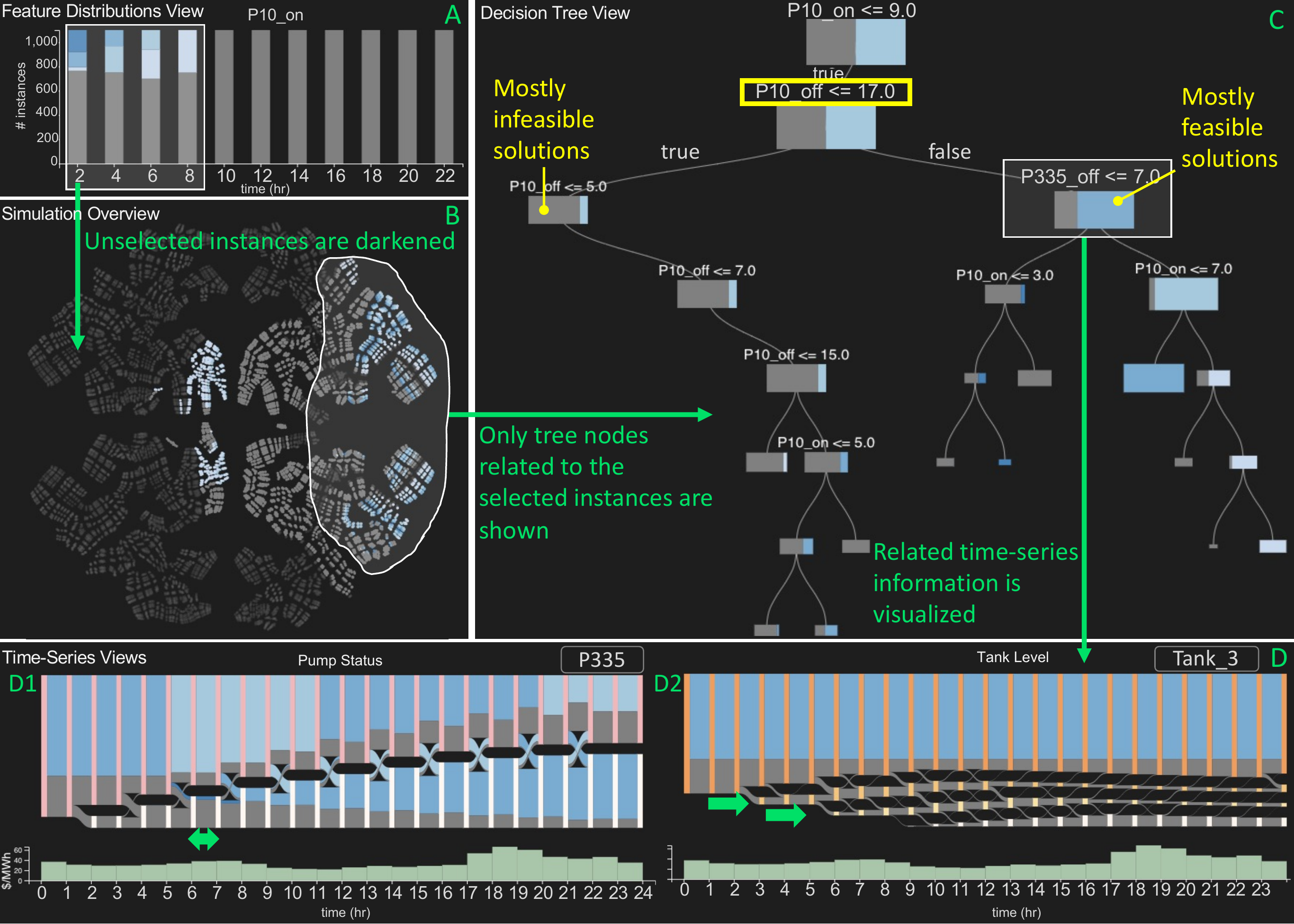}
    \caption{Interactive analysis performed for the Case Study. 
    }
    \vspace{-3pt}
    \label{fig:case1}
\end{figure}

We demonstrate an analysis case of the simulations of the NET 3 model (refer to \autoref{sec:data} for the details of the data) while following the required analysis processes listed in \autoref{sec:DRs}.
We start by looking at the overview of the results (\textbf{DR1}). As shown in~\autoref{fig:teaser},
there is a much larger proportion of infeasible instances in the simulations.
From \autoref{fig:teaser}A, 
we notice that when \texttt{P10\_on}\,$\geq$\,10, the simulation is always infeasible.
From \autoref{fig:teaser}B, we see that feasible solutions are concentrated in two areas, as annotated with the orange boxes.

To examine the range containing feasible solutions in detail (\textbf{DR2}), as shown in \autoref{fig:case1}A, we perform brushing to select the area of \texttt{P10\_on}\,$<$\,10.
From the updated simulation overview (\autoref{fig:case1}B), we can see that the selected instances are placed only at the right side of the view. 
Within these instances, to identify the differences between the infeasible instances and the instances having similar input features but resulting in low energy costs, we perform a lasso selection, as shown in \autoref{fig:case1}B.
Identifying such differences could provide useful insights for further optimization.

Based on the selection, the decision tree view promptly show rules related to the selected instances (\textbf{DR3}).  
From the result shown in \autoref{fig:case1}C, we notice that the rule \texttt{P10\_off}\,$\leq$\,17 (annotated by yellow) strongly separates the feasible and infeasible solutions as its left and right children mainly consist of infeasible and feasible solutions, respectively. 
We further look at the right child of the rule \texttt{P10\_off}\,$\leq$\,17.
This child's rule (i.e., \texttt{P335\_off}\,$\leq$\,7) also seems to clearly separate feasible and infeasible solutions. 
By combining the rules we observed so far, we can say that to achieve low energy costs, \texttt{P10\_on} should be smaller than or equals to 9, \texttt{P10\_off} should be larger than 17, and \texttt{P335\_off} should be larger than 7.

We can further understand the observed rules by using the time-series views (\textbf{DR4}).
Here, as an example, we review the information related to \texttt{P335\_off}\,$\leq$\,7.
From the pump statuses shown in 
\autoref{fig:case1}D1, when \texttt{P335}'s status becomes `off' around hour six and seven (annotated with the green arrow), many simulation instances are categorized as infeasible solutions.
This verifies the rule extracted from the decision tree view.
Also, by looking at the tank levels of \texttt{Tank\_3} shown in \autoref{fig:case1}D2, we observe that infeasible solutions tend to shift tank levels from `high' to lower levels (i.e., `medium high', `medium low', and `low'). 
Moreover, we can see that these shifts tend to more occur before or around hour seven, as indicated with the green arrows.
We conclude that because of the early shutoff of \texttt{P335}, the tank levels become lower at an early stage of the simulation and consequently these simulations violate the constraints on tank levels.

%% file: text_files/6evaluation.tex
\vspace{-2pt}
\section{Expert Review}
\vspace{-2pt}

Our visual analytics system was reviewed by three WDS optimization experts (E1, E2, and E3) to evaluate its usefulness and identify possible improvements for future study.
During the interviews conducted using a video conference setup, we first explained the interface of the system and the purpose of use of each view; then, we presented a live demonstration of the case study.
After the presentation,\,the\,experts\,were\,asked\,to\,provide\,opinions\,for\,a\,list\,of\,questions.

Overall, all three experts expressed that the system is a useful tool to analyze the WDS simulation data.
All of them found the feature distributions view intuitive and informative.
Both E1 and E2 appreciated the use of t-SNE in the simulation overview, saying that it reveals the reliability of one potential solution by viewing its ``surrounding'' solutions (i.e., solutions with similar input features).
E3 also found the decision tree view intuitive for understanding the separability.
As for the actionable insights generated from the system, E1 said that the findings extracted from the system matched their intuition of the WDS, which would increase their confidence in use of the system.
E3 also found that the functionality of filtering to see detailed information and the variability of visualization methods are helpful to generate insights.

On the difficulty of learning to use the tool, all experts confirmed that there should be a training process before using the interface in practice. 
E2 said that, at the beginning of the interview, they did not fully understand the usage and functionality of the system, but once going through the live demonstration of the case study, they clearly understood how to use the system for practical analyses and felt the system design is intuitive.
All experts found that the Sankey diagrams in the time-series views are the least useful visualization.
E1 and E3 found the Sankey diagrams hard to understand without careful explanation, 
while E2 thought that visualizing links used to represent changes in statuses are not necessary.
Also, as a current limitation of the system, E3 mentioned that they wanted to know the breakdown of the violated constrains by infeasible solutions.

%% file: text_files/7discussions.tex
\vspace{-1pt}
\section{Discussion}
\vspace{-1pt}

\textbf{Strengths and limitations.}
From the experts' feedback, the most significant strength of this work is the ability to gain useful and intuitive insights from a large set of simulation data by incorporating computational analysis methods and visualizations.
The simulation overview helps the qualification of an optimization solution's sensitivity, whereas the decision tree view identifies the set of rules that can be used to inform further optimization.

However, limitations still exist in the current system, indicating future improvements.
Firstly, as the Sankey diagrams used in the time-series views are not intuitive enough for the experts, we should consider using simpler visualizations in the future.
Second, as pointed out by E3, some experts would prefer to analyze the simulations with more detailed information. 
To support such requirements, we plan to extend our system to enable flexible  customization of visual encodings. 
As multiple selections are often needed to reach actionable insights, keeping analysis records (i.e., provenance~\cite{ragan2016characterizing}) and summarizing the applied filtering could be beneficial to achieve more efficient analyses.
Also, we can expect that when analyzing larger scale of WDSs (e.g., consisting of 20 pumps) in the future, our system would suffer from the issue of visual scalability, especially in the feature distributions view.
To address this issue, we can utilize the rules obtained by a decision tree to sort and filter input features based on their influence on simulation outcomes.
Finally, comparisons against other choices of visualizations and ML methods are needed to further evaluate the system usability.

\noindent
\textbf{Extensibility.}
Although our system is designed to compare multiple simulations of WDSs, the approaches taken in this work are applicable to other simulation-based optimization problems, which can be found in many real-world applications~\cite{Fu1994,Carson1997,Xu2015}.
By showing multiple aspects of the input space and its relationship to the simulation outcomes,
the visual analytics workflow can effectively provide actionable insights for better optimization. 
Similar to the WDS simulations, multiple runs of these simulations also generate large data consisting of simulation inputs and outcomes. 
And the designs used in the feature distributions view, simulation overview, and decision tree view can be directly applied to other types of simulations to effectively reveal important relationships between simulation inputs and outcomes. 
From these relationships, insights directly usable for the next step of optimization design can be extracted, as we demonstrate in the case study.

%% file: text_files/8conclusion.tex
\vspace{-2pt}
\section{Conclusion}
\vspace{-1pt}
We have presented a visual analytic system that assists in relating multiple simulations' inputs and outcomes to gain actionable insights to optimize water distribution systems (WDSs).
The system incorporates interpretable machine learning to extract general rules underlying complex WDS simulations. 
Our approach reveals general and detailed patterns of simulation runs from multiple aspects. 
This approach is applicable to many other simulations producing a pair of inputs and outcomes.
Thus, our work provides tangible contributions to a wide range of optimization studies using simulations.